\documentclass[aps,prb,twocolumn,superscriptaddress]{revtex4}

\usepackage{graphicx}
\usepackage{dcolumn}
\usepackage{bm}
\usepackage{amsmath, amssymb}
\usepackage{xcolor}

\begin{document}

\title{Triggering phase-coherent spin packets by pulsed electrical spin injection across an Fe/GaAs Schottky barrier}

\author{L. R. Schreiber}
\affiliation{2nd Institute of Physics and JARA-FIT, RWTH Aachen University, 52074 Aachen, Germany}
\affiliation{JARA-FIT Institute for Quantum Information, Forschungszentrum J\"ulich GmbH and RWTH Aachen University, 52074 Aachen, Germany}
\author{C. Schwark}
\author{G. G\"untherodt}
\affiliation{2nd Institute of Physics and JARA-FIT, RWTH Aachen University, 52074 Aachen, Germany}
\author{M. Lepsa}
\affiliation{Peter Gr\"unberg Institute (PGI-10), Forschungszentrum J\"ulich GmbH, 52425 J\"ulich,
Germany}
\author{C. Adelmann}
\affiliation{Department of Chemical Engineering and Material
Science, University of Minnesota, Minneapolis 55455, USA}
\affiliation{IMEC, 3001 Leuven, Belgium}
\author{C. J. Palmstr{\o}m}
\affiliation{Department of Chemical Engineering and Material
Science, University of Minnesota, Minneapolis 55455, USA}
\affiliation{Departments of Electrical and Computer Engineering and
Materials, University of California, Santa Barbara, CA 93106, USA}
\author{X. Lou}
\author{P. A. Crowell}
\affiliation{School of Physics and Astronomy, University of
Minnesota, Minneapolis, Minnesota 55455, USA}
\author {B. Beschoten}
\email[]{bernd.beschoten@physik.rwth-aachen.de}
\affiliation{2nd Institute of Physics and JARA-FIT, RWTH Aachen University, 52074 Aachen, Germany}


\begin{abstract}
The precise control of spins in semiconductor spintronic devices
requires electrical means for generating spin packets with a
well-defined initial phase. We demonstrate a pulsed electrical
scheme that triggers the spin ensemble phase in a similar way as circularly-polarized optical pulses are generating phase
coherent spin packets. Here, we use fast current pulses to initialize phase coherent spin packets,
which are injected across an Fe/GaAs Schottky barrier into $n$-GaAs. By means of time-resolved
Faraday rotation, we demonstrate phase coherence by the observation of multiple Larmor precession
cycles for current pulse widths down to 500 ps at 17 K. We show that the current pulses are
broadened by the charging and discharging time of the Schottky barrier. At high frequencies, the
observable spin coherence is limited only by the finite band width of the current pulses, which is on
the order of 2 GHz. These results therefore demonstrate that all-electrical injection and phase
control of electron spin packets at microwave frequencies is possible in metallic-ferromagnet/semiconductor heterostructures.

\end{abstract}


\maketitle

\section{Introduction}
The preparation and phase-controlled manipulation of coherent single spin states or spin ensembles is fundamental for spintronic devices \cite{Hanson08, Awschalom}.
Devices based on electron spin ensembles requires for spin coherence an initial triggering of the phase of all the individual spins,
which results in a macroscopic phase of the ensemble. Such a phase
triggering can easily be obtained by circularly polarized ultrafast
laser pulses, which are typically shorter than one ps.\cite{Kikkawa98,Kuhlen2014Sep} By impulsive laser excitation, all spins of the ensemble are oriented in the same direction, i.e. they are created with the same initial phase. Spin precession of the ensemble can be monitored by time-resolved
magneto-optical probes as the spin precession time is usually orders
of magnitude longer than the laser pulse width.  Along with other
techniques, these time-resolved all-optical methods have been used
to detect spin dephasing times\cite{Kikkawa98, Schreiber07, Schreiber07a, Schmalbuch2010Dec}, strain-induced spin precession \cite{Kato03, Crooker05} and phase-sensitive spin manipulation in lateral devices
\cite{Kato03,Kuhlen2012Oct,Stepanov2014Feb}.

Spin precession can also be observed in dc transport experiments
\cite{Crooker05Science, Kato05APL, Lou07, Appelbaum07, Huang07,
Li08}. In spin injection devices, for example, electron spins are
injected from a ferromagnetic source into a semiconductor
\cite{Ohno99Nature, Fiederling99, Zhu01, Hanbicki02, Hanbicki03, Jiang05, Adelmann05, Kotissek2007Dec, Truong2009Apr, Asshoff2009Nov, Li2011Mar, Hanbicki2012Feb}. Their initial spin orientation near the ferromagnet/semiconductor interface is defined by the magnetization
direction of the ferromagnet. Individual spins start to precess in a
transverse magnetic field. This results in a rapid depolarisation of
the steady-state spin polarisation (the Hanle effect), because
spins are injected continuously in the time domain. The precessional
phase is preserved partially when there is a well-defined transit
time between the source and the detector \cite{Crooker05Science,
Appelbaum07}. This has been achieved in Si by spin-polarized hot electron injection and detection techniques operated in a drift-dominated regime, which allowed for multiple spin precessions\cite{Appelbaum07, Huang07},
while only very few precessions could be seen in GaAs-based devices\cite{Crooker05Science, Kato05APL}. On the other hand, pulsed electrical spin injection has been reported~\cite{Truong2009Apr, Asshoff2009Nov}, but no spin precession was observed.  Despite recent progress
in realizing all-electrical spintronic devices, electrical phase
triggering is missing.

Here, we use fast current pulses to trigger the ensemble phase of
electrically generated spin packets during spin injection from a
ferromagnetic source into a III-V semiconductor. Coherent precession
of the spin packets is probed by time-resolved Faraday rotation. Our
device consists of a highly doped Schottky tunnel barrier formed
between an epitaxial iron (Fe) and a (100) oriented $n$-GaAs layer. We chose
this device design for three reasons: (I) the Schottky barrier
profile guarantees large spin injection efficiencies
\cite{Hanbicki02, Adelmann05, Schmidt00, Rashba00}, (II) the
$n$-GaAs layer is Si doped with carrier densities near the
metal-insulator transition ($n=2-4 \times 10^{16}$ cm$^{-3}$) which
provides long spin dephasing times $T_2^*$ for detection \cite{Kikkawa98,
Kikkawa99, Dzhioev02} and (III) the Fe injector has a
two-fold magnetic in-plane anisotropy \cite{Crooker07}, which allows
for a non-collinear alignment between the external magnetic field
direction and the magnetization direction of the Fe layer and thus
the spin direction of the injected spin packets. This non-collinear
alignment is needed to induce Larmor precession of the spin
ensemble. We observe spin precession of the electrically injected
spin packets for current pulse widths down to 500 ps. The net
magnetization of the spin packet diminishes with increasing magnetic
field. We link this decrease to the high-frequency properties of the
Schottky barrier. Its charging and discharging leads to a broadening
of the current pulses and hence temporal broadening of the spin
packet as well as phase smearing during spin precession.
We introduce a model for ultrafast electrical spin injection
and extract a Schottky barrier time constant from our Faraday
rotation data of 8~$ \pm ~2$~ns, which is confirmed by independent
high-frequency electrical characterization of our spin device.

\section{Experiment}
Our measurement setup and sample geometry are depicted in Fig. 1a.
The sample consists of an Al-capped 3.5-nm thick, epitaxially grown
Fe(001) layer on $n$-doped Si:GaAs(001). The doping
concentration of the 15-nm thick $n^+$-GaAs layer starting at the
Schottky contact is $5 \times 10^{18}$~cm$^{-3}$ followed by a 15~nm
$n^+/n$ transition layer with a doping gradient, a 5-$\mu$m thick
bulk layer with doping concentration $2 \times 10^{16}$~cm$^{-3}$
and a highly doped ($\sim 1 \times 10^{18}$~cm$^{-3}$) GaAs
substrate (layer stack details in Fig. 3c). The sample mesa with 650~$\mu$m radius is etched down to the substrate. The $T_2^*$ of the substrate is smaller than 1~ns.
The magnetic easy axis of the Fe layer is oriented along the GaAs [011] ($\pm x$ direction). Comparison of electrical and all-optical Hanle measurements indicates a spin injection efficiency into the bulk $n$-GaAs layer of $\sim 7\%$ for a wide bias range.
The differential resistance of the layer stack and the magnetic characterization of the Fe layer is shown in Appendix A.

Samples are mounted in a magneto-optical cryostat kept at 17~K with a magnetic field $B_z$ oriented along the $\pm z$ direction.
For time-resolved electrical spin injection, a voltage pulse train
(amplitude 1.8~V) from a pulse generator (65~ps rise and fall time)
is applied via a bias-tee to the sample, which is placed on a
coplanar waveguide within a magneto-optical cryostat. Linearly polarized laser pulses at normal incidence to the sample plane and phase-locked to the electrical pulses monitor the $\pm y$ component of spins injected in the GaAs by detecting the Faraday
rotation angle $\theta_F$. The linearly polarized laser
pulses ($P=200$~$\mu$W with a focus diameter $\approx 50~\mu$m on
the sample) are generated by a picosecond Ti-sapphire laser with a stabilized
repetition frequency of 80~MHz. They are phase-locked to the voltage
pulses and can be delayed by a time $\Delta t$ up to 125~ns with a variable phase shifter
with ps-resolution. The laser energy 1.508~eV is tuned to just below
the band gap of the GaAs. The repetition interval of the pump and
probe pulses can be altered from 12.5~ns to 125~ns by an optical
pulse selector and the full width at half maximum $\Delta w$ of the
voltage pulses can be varied from 100~ps to 10~ns. Both pump and
probe pulses are intensity-modulated by 50~kHz and 820~Hz,
respectively, in order to extract the pump induced $\theta_F$ signal by a
dual lock-in technique.

\begin{figure}[tb]
\centering
\includegraphics[draft=false,width=0.8\linewidth]{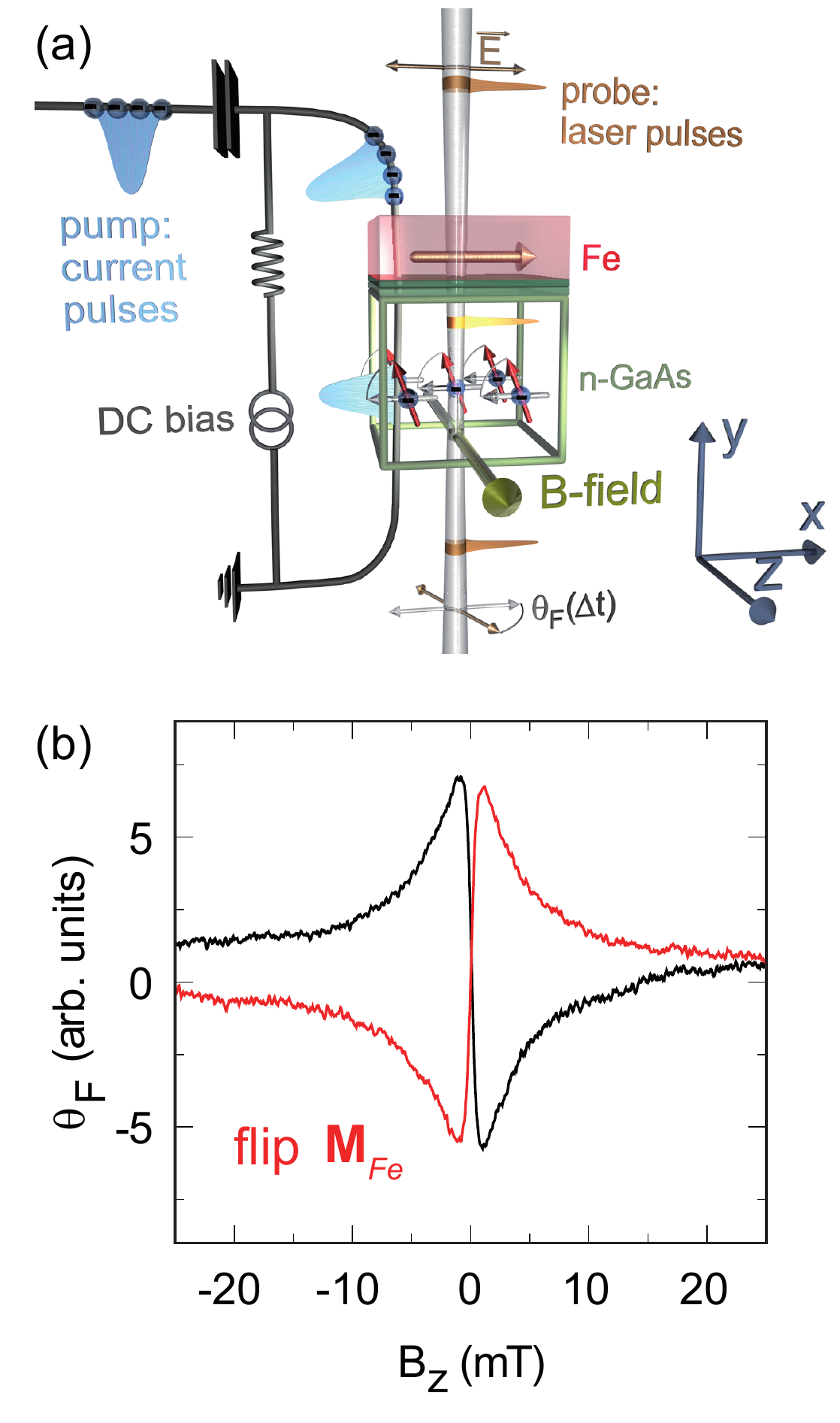}
\caption[Fig01]{ \textbf{Electrical pump and optical
probe setup and test by continuous electrical injection.}  (a) Schematic of the electrical pump and optical probe
experiment of spins injected from Fe into $n$-GaAs. (b)
Faraday rotation $\theta_F$ (Hanle depolarisation) of the dc
current along the $y$ direction as a function of the external
transverse magnetic field $B_z$ at constant bias before (black line)
and after (red line) flipping the Fe magnetization
$\mathbf{M}_{Fe}$.}
\label{f1}
\end{figure}

\section{Results}
\subsection{Static spin injection}
We first use static measurements of the Faraday rotation to
demonstrate electrical spin injection in our devices (Fig. 1b). The
sample is reverse biased, i.e. positive voltage probe on GaAs, and spins are probed near the fundamental band gap of GaAs. At $B_z=0$~T, spins are injected parallel to the
easy axis direction of the Fe layer yielding $\theta_F=0$. At small
magnetic fields $B_z$, spins start to precess towards the
$y$-direction yielding $\theta_F\neq 0$. $\theta_F$ is a direct
measure of the resulting net spin component $S_y$. Changing the sign
of $B_z$ inverts the direction of the spin precession which results
in a sign reversal of $\theta_F$. As expected
\cite{Crooker05Science}, the direction of spin precession also
inverts when the magnetization direction of the Fe layer is reversed
(see red curve in Fig. 1b). $\theta_F$ approaches zero at large
fields, since the continuously injected spins dephase due to Larmor
precession causing strong Hanle depolarisation.

\begin{figure}[tb]
\centering
\includegraphics[draft=false,keepaspectratio=true,clip,width=0.9\linewidth]{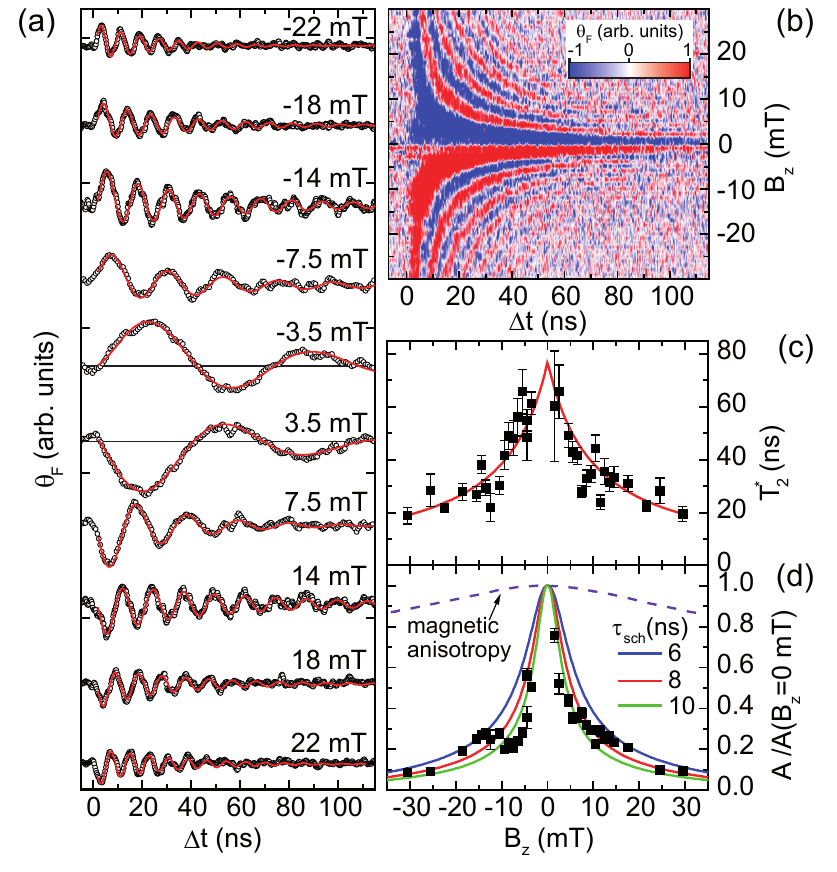}
\caption[Fig02]{\textbf{Pulsed electrical spin
injection.} (a) Time evolution of the Faraday rotation
$\theta_F$ versus pump-probe delay $\Delta t$ for various magnetic
fields $B_z$ with vertical offsets for clarity. Red lines are fits
to the data for $\Delta t
> \Delta w = 2$~ns. (b) False color plot of $\theta_F$ versus $\Delta t$
and $B_z$. (c) Fitted spin dephasing time $T_2^*(B_z)$ and
(d) normalized oscillation amplitudes $A$ versus $B_z$ field.
The error bars include the least-squares fit errors only. The red
solid line in (c) is a least-squares fit to the data and
solid lines in (d) are simulations for different effective charge and discharge times $\tau_{sch}$ of the Schottky barrier. The dashed line is the expected decrease of $A$ due to
the magnetic anisotropy of the Fe injector layer.
}
\label{f2}
\end{figure}

\subsection{Time-resolved spin injection}
For time-resolved spin injection experiments, we now apply voltage
pulses with a full width at half maximum of $\Delta w = 2$~ns and a
repetition time of $T_{rep} = 125$~ns with $T_{rep}>T_2^*$. The
corresponding time-resolved Faraday rotation data are shown in Figs.
2a and 2b at various magnetic fields. Most strikingly, we clearly observe
Larmor precessions of the injected spin packets demonstrating that
the voltage pulses trigger the macro-phase of the spin packets. It
is apparent that the amplitude of $\theta_F$ is diminished with
increasing $|B_z|$. We note that the oscillations in $\theta_F$ are
not symmetric about the zero base line (see black lines in Fig. 2a
as guides to the eye). For quantitative analysis we use

\begin{eqnarray}\label{eq:simplebk}
    \nonumber \theta_F(\Delta t, B_z) &\propto& M_y(\Delta t, B_z)\\
    \nonumber &=& A(B_z) \exp\left(-\frac{\Delta t}{T_2^*(B_z)}\right) \sin(\omega_L \Delta
 t + \phi)\\
    &+& A_{bg}(B_z) \exp\left(-\frac{\Delta t}{\tau_{bg}}\right),
\end{eqnarray}

with $\omega_L = g \mu_B B / \hbar$, where $g$, $\mu_B$ and $\hbar$ denote the effective electron $g$ factor, the Bohr magneton and the reduced Planck constant and $\phi$ being a phase factor. The second term accounts for the non-oscillatory time
dependent background with a lifetime $\tau_{bg}$ and an amplitude $A_{bg}$  (The magnetic field dependence of $A_{bg}$ is shown in the Supplemental Material \cite{SupMat}). The least-squares fits to the experimental
data are shown in Fig. 2a as red curves. We determine a field
independent $\tau_{bg} = 8 \pm 2$~ns and deduce $|g| = 0.42 \pm
0.02$ from $\omega_L$ as expected given that the spin precession is
detected in the bulk $n$-GaAs layer \cite{Kikkawa98}. The extracted
spin dephasing times $T_2^*(B_z)$ and amplitudes $A(B_z)$ are
plotted in Figs. 2c and 2d, respectively. The longest $T_2^*(B_z)$
values, which exceed 65~ns, are obtained at small magnetic fields.
The observed 1/B dependence of $T_2^*(B_z)$ (see red line in Fig.
2c), which indicates inhomogeneous dephasing of the spin packet, is
consistent with results obtained from all-optical time-resolved
experiments on bulk samples with similar doping concentration
\cite{Kikkawa98}. On the other hand, the strong decrease of $A(B_z)$
with magnetic field (Fig. 2d) has not previously been observed in
all-optical experiments. Note that spin precession is barely
visible for magnetic fields above 30~mT.

The $A(B_z)$ dependence might be caused by the $B_z$ field acting on
the direction of the magnetization $\mathbf{M}_{Fe}$ of the Fe
injector. Increasing $B_z$ rotates $\mathbf{M}_{Fe}$ away from the
easy ($x$-direction) towards the hard axis ($z$ direction) of the Fe
layer. This rotation diminishes the $x$ component of the
magnetization vector of the injected spin packet, which would result
in a decrease of $A(B_z)$. We calculated this dependence (see dashed
line in Fig. 2d) for a macrospin $\mathbf{M}_{Fe}$ using in-plane
magnetometry data from the Fe layer (see Fig. 6). The
resulting decrease is, however, too small to explain our $A(B_z)$ dependence.

To summarize, there are two striking observations in our
time-resolved electrical spin injection experiments: (I) the strong
decrease of the Faraday rotation amplitude $A(B_z)$ and (II) the
non-oscillatory background in $\theta_F(\Delta t)$ with a field
independent time constant $\tau_{bg} = 8 \pm 2$~ns. As both have not
been observed in time-resolved all-optical experiments, it is
suggestive to link these properties to the dynamics of the
electrical spin injection process.

In our time-resolved experiment, electron spin packets are injected
across a Schottky barrier by short voltage pulses. The depletion
layer at the barrier acts like a capacitance. When a voltage pulse
is transmitted through the barrier, the capacitance will be charged
and subsequently discharged. For studying the effect of the charging
and discharging on the spin injection process, we performed
high-frequency (HF) electrical characterization of our devices.

\begin{figure}[tb]
\centering
\includegraphics[draft=false,keepaspectratio=true,clip,width=0.9\linewidth]{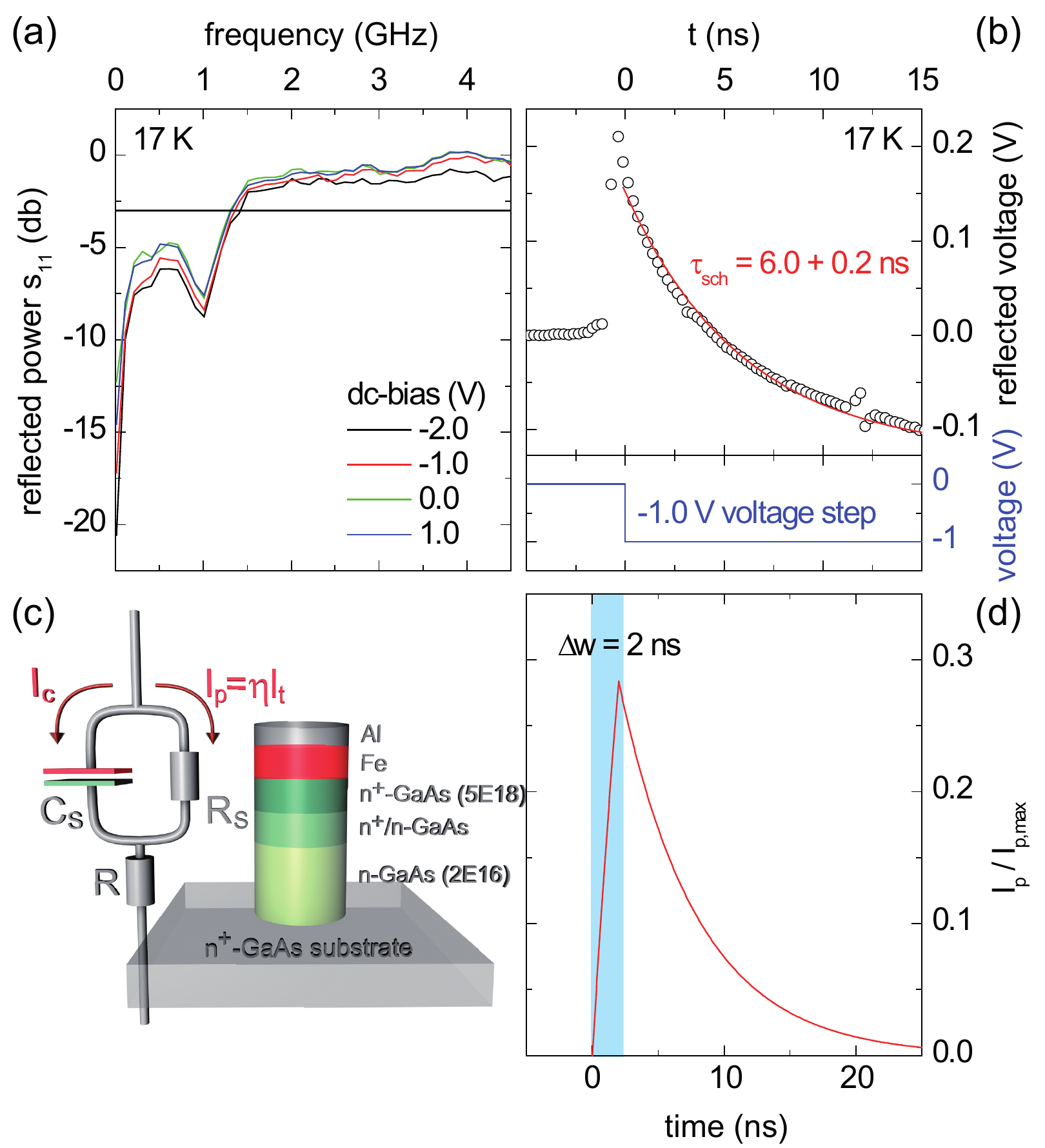}
\caption[Fig03]{\textbf{Electrical high frequency
characterization.} (a) Reflected power ($S_{11}$) for various
dc-bias operating points obtained from vector network analysis.
(b) Voltage reflected from the sample after applying a
voltage step function (-1.0V) with least-squares exponential fit
(red solid line). (c) Sample structure and simple equivalent
network of the sample. (d) Simulation (red line) of the
evolution of the spin-polarized tunnel current through the Schottky
barrier triggered by a 2~ns long current pulse (light blue).
}
\label{f3}
\end{figure}

\subsection{High-frequency sample characteristic}
The HF bandwidth of the sample is deduced from the reflected
electrical power $S_{11}$ by vector network analysis as shown in Fig.
3a. More than half of the electrical power ($S_{11}> 3$~dB) is
reflected from the device for frequencies above $\sim 1.5$~GHz. This
bandwidth is independent of the operating point over a wide dc-bias
range from -2.0~V (reverse biased Schottky contact) to 1.0~V and
allows the sample to absorb voltage pulses of width $\Delta w
\gtrsim 500$~ps.

Furthermore, the time evolution of the voltage drop
at the Schottky barrier, i.e., its charging and discharging, can
directly be determined by time-domain reflectometry (TDR). To analyse the charging dynamics of the Schottky capacitance, we apply a voltage step to the sample with an amplitude of -1~V and a rise time of 100~ps. The time-evolution
of the reflected voltage step is shown in Fig. 3b. Note that there
is a significant temporal broadening of the voltage step. We obtain
a similar time constant for the discharging behaviour (not shown).
Any impedance mismatch along the 50~$\Omega$ transmission
line can be detected by measuring the time evolution of the
reflected voltage. A real impedance above 50~$\Omega$ yields a
reflected step function with negative amplitude. If the transmission line is
terminated by a capacitance, the time evolution of the voltage drop
during charging of the capacitance equals the time
dependence of the reflected voltage. Note that even after 15~ns the voltage pulse is not fully absorbed by the sample, i.e. about 10~\% of its amplitude is still being
reflected.  As long as the pulse is applied the absolute amplitude of the reflected voltage will rise towards saturation, which is reached at full charging up of the capacitance (Further information is provided in the Supplemental Material \cite{SupMat}).

To further link the HF dynamics of the Schottky barrier to the
pulsed electrical spin injection process, we depict a simple
equivalent network of the sample in Fig. 3c. In the reverse-bias
regime, the Schottky contact can be modeled by a Schottky
capacitance $C_s$ and a parallel tunnel-resistance $R_s$. The
underlying $n$-GaAs detection layer is represented by a resistance
$R$ in series. We assume the displacement current $I_c$ to be
unpolarized, while the tunneling current $I_t$ carries the spin
polarized electrons. The spin current $I_p = \eta I_t$ is given by
the spin injection efficiency $\eta$. The charging and discharging
of the Schottky capacitance is thus directly mapped to the temporal
evolution of the spin current. $I_p$ increases after the voltage
pulse is turned on, whereas it decreases after the pulse is turned
off after time $\Delta w$, i.e. during the discharge of $C_s$. If
$C_s$, $R_s$ and $\eta$ are approximately bias-independent, the
increase and decrease of $I_p$ is single-exponential

\begin{widetext}
\begin{equation}\label{eq:Ip}
     I_{p}(t)=I_{p,dc} \times \left\{
\begin{array}{c|l} 1-d \exp\left(-\frac{t}{\tau_{sch}}\right) & 0\leq
t<\Delta w \\
\left[ \exp \left(\frac{\Delta w}{\tau_{sch}} \right)-d \right] \exp
\left(-\frac{t}{\tau_{sch}} \right) & \Delta w \leq t < T_{rep}
   \end{array}
     \right.
\end{equation}
\end{widetext}

\noindent and determined by the effective charge and discharge time $\tau_{sch}$ of the Schottky barrier as illustrated in Fig. 3d
for a pulse width of $\Delta w=2$~ns and $\tau_{sch}=6$~ns. The
constant $d=\left[\exp \left( \frac{\Delta w-T_{rep}}{\tau_{sch}}
\right)-1 \right] / \left[\exp \left(- \frac{T_{rep}}{\tau_{sch}}
\right)-1\right]$ is given by the boundary condition
$I_p(0)=I_p(T_{rep})$.

It is important to emphasize that the temporal width of the
electrically injected spin packet is determined by $\tau_{sch}$.
This temporal broadening becomes particularly important when
individual spins start to precess in the external magnetic field at
all times during the spin pulse. The retardation of spin precession
results in spin dephasing of the spin packet. This phase "smearing"
leads to a decrease of the net magnetization. Its temporal evolution
can be estimated by

\begin{equation}
 M_y(B_z,\Delta t) = \int\limits_0^{\Delta t} dt\ r_S(t)  M_0(\Delta t -
 t),
\end{equation}

\begin{figure}[tb]
\centering
\includegraphics[draft=false,keepaspectratio=true,clip,width=0.9\linewidth]{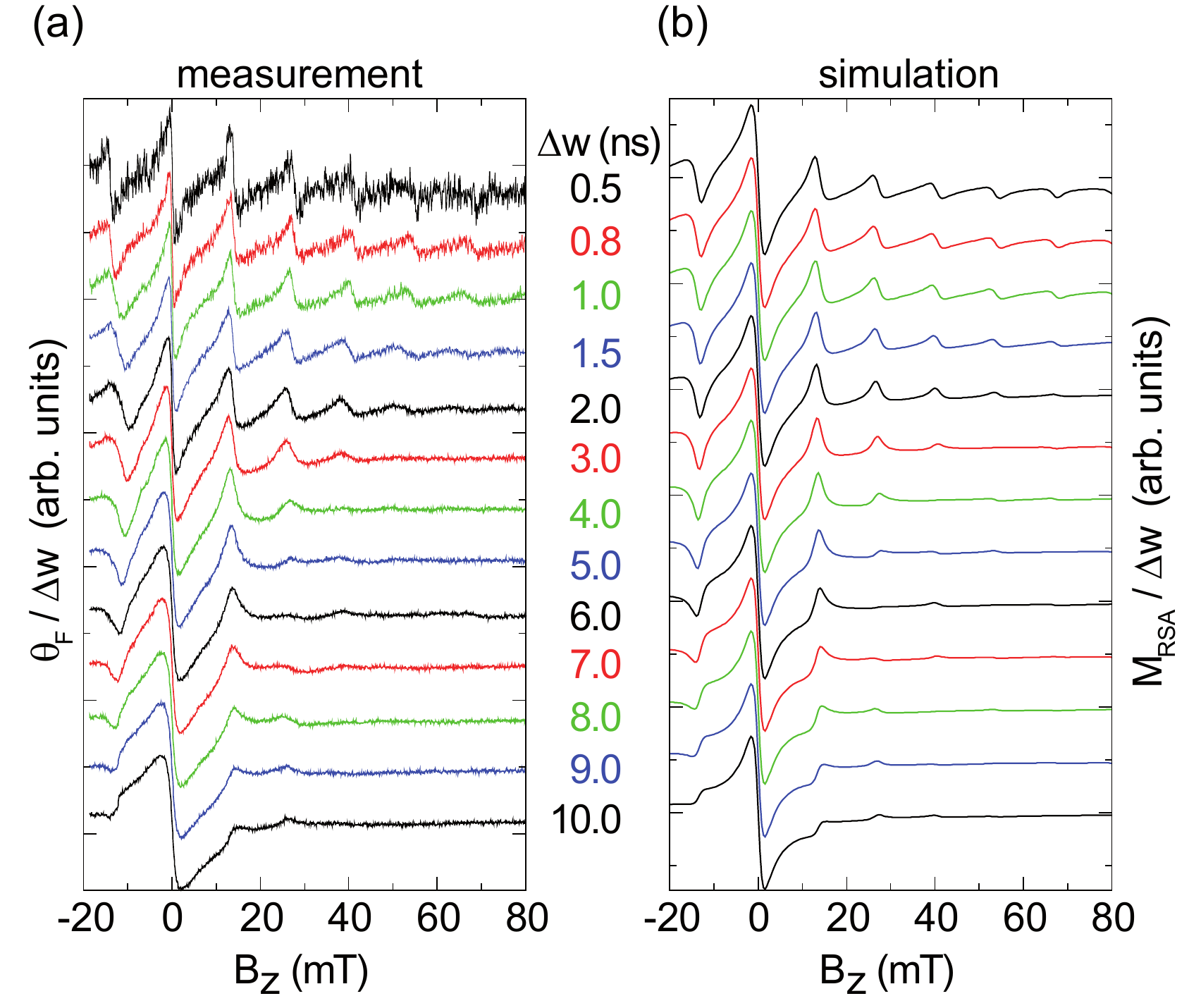}
\caption[Fig04]{\textbf{Resonant spin amplification for
various voltage pulse widths $\Delta w$.} (a) Measured
Faraday rotation $\theta_F$ normalized to $\Delta w$ versus magnetic
field $B_z$ at $T_{rep}=12.5$~ns and fixed $\Delta t$ with vertical
offsets for clarity. (b) Simulations applying Eqs. 2 and 4
with $r_s(t)$ from Fig. 3d and $\tau_{sch}=6$~ns, $T_2^*=18$~ns.
}
\label{f4}
\end{figure}

\noindent where $r_S(t)=I_p(t)/a$ is the spin injection rate with
the active sample area $a$ and where $M_0$ is given by an
exponentially damped single spin Larmor precession. The integral can
be solved analytically \cite{SupMat} and results in a
form as given qualitatively by Eq. 1 describing the dynamics of the
injected spin packets, assuming $I_p(0)=0$, i.e., $d=1$. Note that
the non-precessing background signal of $\theta_F$ (see Fig. 2a)
stems from the discharging of the Schottky capacitance, i.e.
$\tau_{sch}=\tau_{bg}$, while $T_2^*$ is not affected by the
integration. This assignment is confirmed by the independent
determination of $\tau_{sch}$ by TDR. The amplitude $A(B_z)$ in Eq.
1 becomes a function of $\omega_L, T_2^*, \tau_{sch}, \Delta w$ and
$r_s$ (see Eqs. S17 and S22 of the Supplemental Material \cite{SupMat}. For simulating
$A(B_z)$, we take the above fitting results from Fig. 2, i.e.
$T_2^*(B_z)$, $\omega_L$, as well as $\Delta w=2$~ns and vary only
$\tau_{sch}$ as a free parameter. The resulting field dependent
amplitudes are plotted in Fig. 2d at various time constants
$\tau_{sch}$. The experimental data are remarkably well reproduced
for the $\tau_{sch}$ values determined by TDR ($\tau_{sch}=6$~ns)
and by the non-oscillatory background of $\theta_F$
($\tau_{sch}=8$~ns). This demonstrates that the charging and
discharging of the Schottky capacitance is the main source of the
amplitude drop in our experiment.

\subsection{Resonant spin amplification}
We now analyse the precession of the spin packets after injection
with voltage pulses of different width $\Delta w$. This can better
be tested as a function of $B$ field instead of in the time domain.
To enhance the signal-to-noise ratio of $\theta_F$, we reduce
$T_{rep}$ to 12.5~ns. As $T_{rep}$ is now shorter than $T_2^*$, spin
packets from subsequent voltage pulses can interfere. We thus enter
the regime of resonant spin amplification (RSA)\cite{Kikkawa98, Kuhlen2014Sep}. The
net RSA magnetization $M_{y,RSA}$ results with Eq. 3 in

\begin{align}
 M_{y,RSA}(B_z,\Delta t) & = M(\Delta t) \\
 & + \sum\limits_{n=1}^{\infty} \int\limits_0^{T_{rep}} dt\ r_S(t) M_0(\Delta t -t + nT_{rep}), \nonumber
\end{align}

\noindent where $M_{RSA}$ and $r_S$ are periodic in $T_{rep}$ and
defined in the time interval $[0, T_{rep})$. Constructive
interference of subsequent spin packets leads to periodic series of
resonances as a function of $B$, if a multiple of
$1/T_{rep}$ equals the Larmor frequency:

\begin{equation}
  z / T_{rep} = \omega_L / (2 \pi),
\end{equation}

\noindent where $z$ is an integer.

Fig. 4a shows RSA scans for $\Delta w$ ranging between 500~ps and
10~ns taken at fixed $\Delta t$ and normalized to $\Delta w$.
Multiple resonances are observed for short $\Delta w\leq2$~ns. The
strong decrease of the resonance amplitudes with the increase of
$|B_z|$ is consistent with the time-domain experiments (see Fig. 2).
The number of resonances, which equals the number of Larmor
precession cycles, subsequently decreases for broader current
pulses. We observe a continuous crossover to the Hanle regime
for the broadest pulses of $\Delta w=10$~ns $ \sim T_{rep}=12.5$~ns,
which is close to the dc-limit of spin injection as shown in Fig.
1b. This crossover strikingly demonstrates the phase triggering by
the current pulses. While pulse-width induced phase smearing is
observed above $\Delta w=1.5$~ns, there are no effects of the pulse
width below 1.5~ns due to the finite $\tau_{sch}$. Remarkably,
pulsed spin injection is possible for $\Delta w$ as short as 500~ps.

The RSA scans are simulated using Eqs. 2 and 4 with
$\tau_{sch}=6$~ns and are depicted in Fig. 4b. The dependence on
$B_z$ as well as the phase "smearing" with increasing pulse width
are well reproduced. Note that even the change of the RSA peak shape
for higher order resonances is reproduced by the simulations,
demonstrating that our model explains all salient features of the
experiment.

\section{Conclusion}
In conclusion, we have shown that fast current pulses can trigger
the macroscopic phase of spin packets electrically injected across
an Fe/GaAs Schottky barrier. Current pulses having a width down to 500~ps trigger a spin imbalance observed as magnetic oscillations matching the effective electron g-factor of GaAs. Charging and discharging of the Schottky barrier yield a temporal broadening of the spin packets resulting in a partial dephasing during spin precession. This partial spin dephasing manifests itself in a characteristic decrease of the oscillation amplitude as a function of the magnetic field and as a non-oscillating exponential decrease of the injected spin-magnetization. Our model fully captures both of these features, which have not appeared when using ultra-fast laser pulses for optical spin orientation, and it predicts that the time constant of the decreasing background is given by the discharging time constant of the Schottky barrier. This time constant independently determined by time-domain reflectometry well matches our observations of the phase smearing of the spin packet. Using a ten time higher frequency of the current pulses, we superimpose injected spin packets in GaAs and enter the regime of resonant spin amplification, which is well-covered by our model as well. Our model predicts that the phase smearing can be significantly suppressed by reduction of the the Schottky capacitance. In this respect spin injection from diluted magnetic semiconductors will be advantageous for realizing
all-electrical coherent spintronic devices of high frequency bandwidth.

\section{Acknowledgments}
Supported by HGF and by the Deutsche Forschungsgemeinschaft (DFG, German Research Foundation) under SPP 1285 (Grant no. 40956248). C.J.P and P.A.C acknowledge funding from the Office of Naval Research, the National Science Foundation (NSF) MRSEC program, the NSF NNIN program, and the University of Minnesota.

\begin{appendix}
\section{Sample characteristics}
\begin{figure}[tb]
\includegraphics{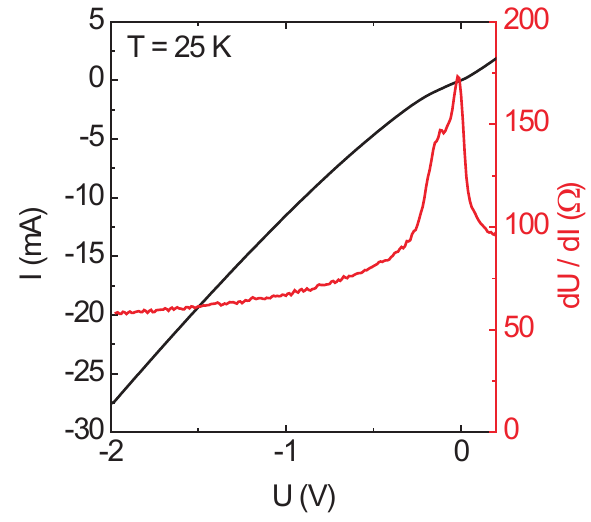}
\caption{\label{fig1} \textbf{IV-characteristics.} dc-current $I$ as a
function of dc-bias $U$ (black line) and differential resistance $dU/dI$
(red line) of the sample with $1300$~$\mu$m diameter mesa at $25$~K. In our
experiment, the Schottky diode is reverse-biased using short $-1.8$~V
voltage pulses.}
\end{figure}

\begin{figure}[tb]
\includegraphics{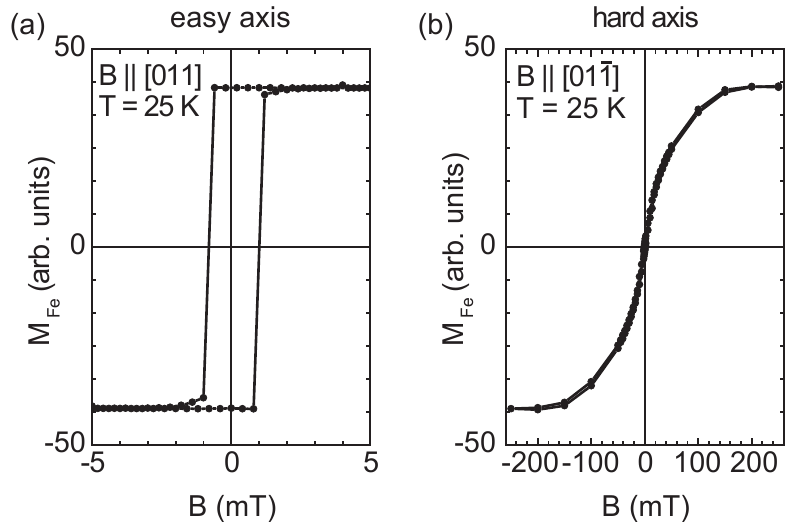}
\caption{\label{fig2} \textbf{Magnetic anisotropy of the Fe injector
layer.} In-plane magnetization of the epitaxial iron injector layer
$M_{Fe}$ as a function of the external magnetic field $B$ as determined
from superconducting quantum interference device (SQUID) measurements at 25
K. The $B$-field is applied parallel to (a) the GaAs [011] and (b) the GaAs
[01$\overline{1}$]  crystal directions. The Fe injector
layer exhibits an in-plane anisotropy. In our experiment, $B$ is applied
nearly parallel to the GaAs [01$\overline{1}$] direction (hard axis).}
\end{figure}

This section provides additional information about the sample used
in our experiment. The I-V characteristics are displayed in Fig.
\ref{fig1}. Magnetometry data of the Fe injector can be found in
Fig. \ref{fig2}

\end{appendix}

\def \dt{\Delta t}
\def \dtt{\Delta t^{\prime}}
\def \tstern{T_{2}^{*}}
\def \ts{\tau_{sch}}
\def \dw{\Delta w}
\def \lam{\omega_{L}}


\renewcommand{\figurename}{Fig.}

\widetext
\pagebreak
\def \dt{\Delta t}
\def \dtt{\Delta t^{\prime}}
\def \tstern{T_{2}^{*}}
\def \ts{\tau_{sch}}
\def \dw{\Delta w}
\def \lam{\omega_{L}}
\begin{center}
	\textbf{\large Supplemental Material: Triggering phase-coherent spin packets by pulsed electrical spin injection across an Fe/GaAs Schottky barrier}
\end{center}

\setcounter{equation}{0}
\setcounter{figure}{0}
\setcounter{table}{0}
\setcounter{section}{0}
\setcounter{page}{1}

\makeatletter
\renewcommand{\theequation}{S\arabic{equation}}
\renewcommand{\thefigure}{S\arabic{figure}}
\renewcommand{\thesection}{S\arabic{section}}
\renewcommand{\thetable}{S\arabic{table}}

\section{Superposition of pulse sequences}
\begin{figure}[tbh]
\includegraphics{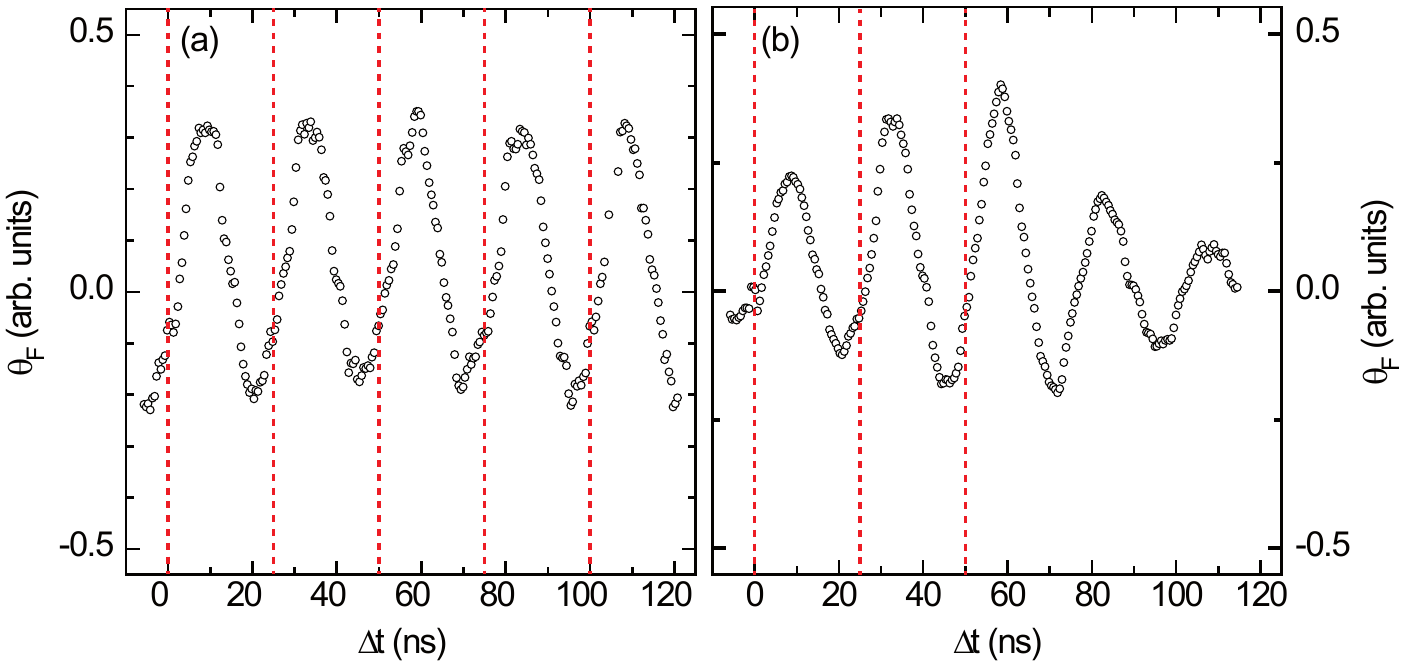}
\caption{\label{figs3} \textbf{Superposition of pulse sequences.} Measured
Faraday rotation $\theta_F$ as a function of the pump-probe delay $\Delta t$. The repetition frequency is set to $T_{rep}=125$~ns, $\Delta w=2$~ns and $B=6.6$~mT. (a) The probe pulse repetition is set to $T_{rep}=125$~ns and excitation pulses are applied at $\Delta t={0, 25, 50, 75, 100}$~ns (marked by red dashed vertical lines). (b) Same as in panel a, but only three excitation pulses are applied at $\Delta t={0, 25, 50}$~ns (marked by red dashed vertical lines).}
\end{figure}

In Fig. 2 of the main article, a repetition frequency of the pump/probe pulses of $T_{rep}=125$~ns is used. At a repetition frequency of $T_{rep}=12.5$~ns (Fig. 4), which is lower than the spin coherence time $\tstern$, the injected spin pulses start to superimpose and we observe resonant spin amplification. Here we consider a repetition period $T_{rep}=125$~ns for the probe laser-pulse. In addition to the electrical pump pulse at $\Delta t=0$, up to four additional current pulses can be applied each delayed by an additional 25~ns, while the repetition of probe pulses is kept at $125$~ns. Hence, we observe the superposition of injected spin packets in the time domain. First, we apply pump pulses with a repetition period of $25$~ns and choose $B=6.6$~mT for the magnetic field, such that the Larmor period equals the pump-pulse period: Sequential spin packets constructively superimpose and we enter the resonant spin amplification regime (Fig. \ref{figs3}a). The period of the $\theta_F$ signal equals the pump repetition frequency as expected. In Fig. \ref{figs3}b, we leave out the last two pump pulses of the sequence within the $125$~ns probe period. Accordingly, we observe the rise of the magnetization due to the constructive superposition of the first three spin-polarized current pulses in the first half of the probe-pulse period, followed by a decrease of the magnetization due to dephasing of the spin packets in the second half of the probe-pulse period.

\section{Model and derivations}
In this section, we derive the fit Eq. 1 of the article from the
ansatz Eq. 3. This calculation yields the expression of the
simulated decrease of the Faraday rotation amplitude $A(B)$ as a
function of the transverse magnetic field $B$ as plotted in Fig. 2d
of the article.

Our model is based on the equivalent circuit diagram (Fig. 3c of the
article), in which the Schottky contact is replaced by a capacitance
$C_s$ and a parallel resistance $R_s$. The spin-polarized current
$I_p(t)$ from the Fe-injector into the semiconductor is only
transmitted by the tunnel current $I_t(t)$ through the Schottky
resistance $R_S$. The displacement current $I_c(t)$ through the
capacitance is assumed to be unpolarized. In order to
determine $I_p(t)$, we use a local model neglecting runtime effects,
since the distance of the elements in the equivalent circuit diagram
is by far smaller than the electric AC wavelength used. From Fig. 1 it
can be deduced that the Schottky contact in the reverse bias regime
($U < 0$~V) is mainly ohmic. Since we use voltage pulses with an
amplitude as high as $-1.8$~V, we can for simplicity neglect the
weak bias-dependence of the tunnel resistance. When an external
negative bias pulse of width $\dw$ is applied to the sample at time
$t=0$, the Schottky capacitance starts to charge. If we assume for
simplicity that the magnitude of the capacitance is constant, the
$I_t(t)$ through the parallel resistor starts to rise exponentially
and approaches the tunnel current $I_{t,dc}$, at which the Schottky
capacitance would be fully charged. When the applied external bias
is switched off at time $t=\dw$, the Schottky capacitance starts to
discharge. This yields an exponential decrease of the voltage
dropping across the parallel resistance $R_S$, and thus the tunnel current
$I_t$ decreases exponentially with the characteristic decay time
denoted $\ts$. If we further assume that the spin injection
efficiency $\eta$ is bias-independent, the polarized current is
$I_p(t)=\eta I_t(t)$. In the case of a dc-bias applied to the
sample, the spin injection rate reaches its maximum value
$I_{p,dc}=\eta I_{t,dc}$. Hence, if a single voltage pulse is
applied starting at time $t=0$, the polarized current is

\begin{equation}\label{eq:polarizedcurrent}
    I_p(t)=I_{p,dc} \times \left\{ \begin{array}{c|l}
                       1-d \exp\left(-\frac{t}{\ts}\right) & 0\leq t<\dw \\
                       \left[ \exp \left(\frac{\dw}{\ts} \right)-d \right] \exp \left(-\frac{t}{\ts} \right) & \dw\leq t<T_{rep}
                     \end{array}
    \right.
\end{equation}

\noindent with the time constant $\ts$ for charging and discharging
the Schottky capacitance. The constant $d$ is determined by the
boundary conditions, i.e. the charging state of the Schottky capacitor, when the next current pulse arrives. For example, for a pulse repetition time $T_{rep}$ much longer than $\ts$, the Schottky capacitor is fully discharged and the spin-polarized current is
$I_p(0)=I_p(T_{rep})=0$ and hence $d=1$. The time-evolution of the
voltage dropping at the sample and thus $\ts$ can be determined
directly from time-domain reflectometry as plotted in Fig. 3b of the
article.

Now, we calculate the effect of the time-dependent polarized current
$I_p(t)$ on the time-evolution of the observed Faraday rotation signal
$\theta_F(\dt,B)$ in a transverse magnetic field $B$. The precession of a
coherent spin packet in a transverse magnetic field is observed by its
magnetization $\mathbf{M}$. Let us start the calculation with the simple case of a purely coherent spin injection, i.e. all spins are injected exactly at the same time, with a magnetization denoted by $\mathbf{M}_0$. This case is relevant for optical spin orientation by an ultra-short laser pulse, which is much shorter than the Larmor precession period of the oriented electron spins. Since the electrically injected spins are pumped perpendicular to the observation direction, which is determined by the probe laser beam, $\theta_F(\dt)$ is
then proportional to $M_0^\bot(\Delta t, B)$

\begin{equation}\label{eq:singlespin}
    M_0^\bot(\Delta t, B) \propto \exp \left( -\frac{\Delta t}{T_2^*} \right) \sin
    \left( \omega_L \Delta t \right),
\end{equation}

\noindent where $\dt$, $T_2^*$, $\omega_L$ denote the pump-probe
delay, the spin dephasing time, the Larmor frequency, respectively. Using a complex $M_0(\Delta t, B)$ with
$M_0^\bot(\Delta t, B) = \Im (M_0(\Delta t, B))$, the calculation
becomes independent of the observation direction:

\begin{equation}\label{eq:complexsingle}
    M_0(\dt, B) \propto \exp \left( -\frac{\Delta t}{T_2^*} \right) \exp
    \left(i \omega_L \Delta t \right).
\end{equation}
\noindent The proportionality factor depends on the number of injected electrons and the magnetic moment of a single electron. It does not depend on the external magnetic field $B$.

Now, we take into account that the spins are injected slowly compared to the Larmor precession frequency. Thus, the first electron spins already precess, when further electrons are injected in the direction given by the static magnetisation of the iron layer. The probe laser measures the total magnetization $M$ induced by the injected spins, by
$\theta_F$:

\begin{eqnarray}\label{eq:netmagnetization}
    \nonumber M(\dt, B) &\propto& \int_0^{\dt} I_p(t) \exp \left( -\frac{\dt -t}{T_2^*} \right) \exp
    \left(i \omega_L (\dt-t) \right)\ dt\\
    \nonumber &\propto& \exp \left( -\frac{\dt}{T_2^*}+i \omega_L \dt \right)\int_0^{\dt} I_p(t) \exp \left(\frac{t}{T_2^*} -i \omega_L t)
    \right)\ dt\\
    &\propto& M_0(\dt, B) \int_0^{\dt} I_p(t) M_0(-t, B)\ dt.
\end{eqnarray}
\noindent Despite its closed form, Eq. \ref{eq:netmagnetization} is
a complex integral over retarded purely coherent spin precessions $M_0(t,B)$,
which is not suitable for data fitting. In the experiment, however,
we observe the precessing net magnetization of the total injected
spin ensemble by the Faraday rotation of the probe beam. In the
following, our goal is to transform Eq. \ref{eq:netmagnetization} to
Eq. 1 of the article used to fit the measured $\theta_F(\dt,B)$
signal.

\noindent We start discussing $M(\dt,B)$ separately during the charging
($0\leq \dt<\dw$) and discharging ($\dw\leq\dt<T_{rep}$) process of the
Schottky contact and define:

\begin{equation}
    M(\dt, B)=\left\{ \begin{array}{c|l}
                       M_{cha}(\dt, B) & 0\leq \dt<\dw \\
                       M_{dis}(\dt, B) & \dw\leq\dt<T_{rep}
                     \end{array}
    \right.
\end{equation}

\noindent It follows with Eq. \ref{eq:polarizedcurrent} and \ref{eq:netmagnetization} (here $d=1$)

\begin{eqnarray}
  \nonumber M_{cha}(\dt, B) &\propto& I_{p,dc} M_0(\dt, B) \int\limits_0^{\dt}\left[1-\exp\left(-\frac{t}{\ts}\right)
  \right]M_0(-t, B)\ dt\\
  \nonumber M_{dis}(\dt, B) &\propto& I_{p,dc} M_0(\dt,B) \int\limits_0^{\dw}\left[1-\exp\left(-\frac{t}{\ts}\right)
  \right]M_0(-t, 0)\ dt+ \\  \nonumber & & + I_{p,dc} M_0(\dt,B) \int\limits_{\dw}^{\dt}\left[\left(\exp\left(\frac{\dw}{\ts}\right)-1 \right)\exp\left(-\frac{t}{\ts}\right)
  \right]M_0(-t, B)\ dt \\
  \label{eq:term1} &=:& M_{dis}^{(1)}(\dt, B) + M_{dis}^{(2)}(\dt, B) \\
\end{eqnarray}

\noindent Note that we replace $I_p(t)$ by the spin injection rate $r_S(t)=I_p(t)/a$ with
the active sample area $a$ in the main article. We absorb here the $B$-independent factor $a$ in the proportionality factor. In the following, we consider only the net magnetization
$M_{dis}(\dt, B)$ during the \emph{discharge} of the capacitance
($\dt\geq\dw$) following the spin injection process for
$0\leq\dt<\dw$, when the external bias is applied. This approach is
sufficient, since only the domain $\dt\in [\dw,T_{rep})$ is used for
the least-square fits in Fig. 2a of the article.

\noindent For compact writing, we define the real constant factor

\begin{equation}\label{factor}
  f=I_{p,dc}\left(\exp\left(\frac{\dw}{\ts}\right)-1 \right),
\end{equation}
\noindent which is independent of $B$ and leave out the explicit $B$-dependence of $M_0(t,B)=M_0(t)$. First, we transform the second summand $M_{dis}^{(2)}(\dt, B)$ into a precessing net magnetization. Therefore, it is useful to introduce a
timescale $\dtt = \dt-\dw$:

\begin{eqnarray}
  \nonumber M_{dis}^{(2)}(\dtt,B) &\propto& f\ M_0(\dtt+\dw) \int\limits_{\dw}^{\dtt+\dw} \exp\left(-\frac{t}{\ts}\right)
  M_0(-t)\ dt \\
  \nonumber  &\propto& f\ M_0(\dtt+\dw) \int\limits_{0}^{\dtt} \exp\left(-\frac{t+\dw}{\ts}\right)
  M_0(-t-\dw)\ dt
\end{eqnarray}

\noindent Making use of the relation
$M_0(-t-\dw)=\exp(\dw/\tstern-i \lam \dw)\ M_0(-t)$ according to
Eq. \ref{eq:complexsingle}, we can further simplify the equation:

\begin{eqnarray}
  \nonumber  M_{dis}^{(2)}(\dtt, B) &\propto& f \exp \left( -\frac{\dw}{\ts}\right) M_0(\dtt)\exp\left(\frac{-\dw}{\tstern}\right) \int\limits_{0}^{\dtt} \exp\left(-\frac{t}{\ts}\right)
  M_0(-t)\exp\left(\frac{\dw}{\tstern}\right)\ dt \\
  \nonumber &\propto& f \exp \left( -\frac{\dw}{\ts}\right) M_0(\dtt) \int\limits_{0}^{\dtt} \exp\left(-\frac{t}{\ts}\right)
  M_0(-t)\ dt \\
&\propto& I_{p,dc} \left(1-\exp\left(-\frac{\dw}{\ts}\right)
\right) M_0(\dtt) \int\limits_{0}^{\dtt} \exp\left(-\frac{t}{\ts}\right)
  M_0(-t)\ dt \label{eq:Mintermediate}
\end{eqnarray}

\noindent $M_{dis}^{(2)}(\dtt, B)$ denotes the total precessing
magnetization at time $\dtt$, if spins are injected with an
exponentially damped injection rate starting at $\dtt=0$. Hence, it
is due to the exponentially damped tail of the polarized current in
Fig. 3d of the main article. In order to simplify $M_{dis}^{(2)}(\dtt, B)$, we briefly neglect the $B$ and $\dtt$-independent real prefactor and focus on solving the integral and write for brevity $\gamma=1/\tstern$ and $\lambda=1/\ts$:

\begin{eqnarray}
  \nonumber  M_{dis}^{(2)}(\dtt, B) &\propto& M_0(\dtt) \int\limits_{0}^{\dtt} \exp\left(-\lambda t\right)
  M_0(-t)\ dt \\
\nonumber &\propto&  \exp\left((i\omega_L-\gamma)\dtt\right) \int\limits_{0}^{\dtt} \exp\left((\gamma-\lambda-i\omega_L)t\right)\ dt \\
\nonumber &\propto& \frac{ \exp\left((i\omega_L-\gamma)\dtt\right)}{\gamma-\lambda-i\omega_L} \left[ \exp((\gamma-\lambda-i\omega_L)\dtt)-1\right]  \\
\nonumber &\propto& \frac{1}{\gamma-\lambda-i\omega_L} \left[ \exp(-\lambda \dtt)- \exp((-\gamma+i \omega_L)\dtt)\right]  \\
\end{eqnarray}
Since we observe the spins by Faraday rotation parallel to the $y$-direction and thus perpendicular to
their original polarization direction in the Fe layer (parallel to $x$-direction), we are interested in
$M_{dis}^{\bot(2)}(\dtt, B)=\Im\left(M_{dis}^{(2)}(\dtt,
B)\right)$, which results in
\begin{eqnarray}
\nonumber M_{dis}^{\bot(2)}(\dtt, B) &\propto& \frac{\omega_L e^{-\lambda \dtt}}{(\gamma-\lambda)^2+\omega_L^2}- \frac{\gamma e^{-\gamma \dtt} \sin(\omega_L \dtt)}{(\gamma-\lambda)^2+\omega_L^2}+\frac{\lambda e^{-\gamma \dtt} \sin(\omega_L \dtt)}{(\gamma-\lambda)^2+\omega_L^2}-\frac{\omega_L e^{-\gamma \dtt} \cos(\omega_L \dtt)}{(\gamma-\lambda)^2+\omega_L^2} \\
\nonumber &\propto& \frac{\omega_L}{(\gamma-\lambda)^2+\omega_L^2} \left[e^{-\lambda \dtt}-e^{-\gamma \dtt}\left(\cos(\omega_L \dtt)+\frac{\gamma-\lambda}{\omega_L} \sin(\omega_L \dtt)\right)\right] \\
\nonumber &\propto& \frac{1/\omega_L}{\Gamma^2+1} \left[e^{-\lambda \dtt}-e^{-\gamma \dtt}\left(\cos(\omega_L \dtt)+\Gamma \sin(\omega_L \dtt)\right)\right].
\end{eqnarray}

\noindent In the last step, we introduced the unit-less constant for brevity:
\begin{equation}\label{eq:charconstant}
    \Gamma = \frac{\gamma-\lambda}{\omega_L}=\frac{1}{\lam} \left(\frac{1}{\tstern}-\frac{1}{\ts}   \right).
\end{equation}

\noindent Adding the real prefactor from Eq. \ref{eq:Mintermediate}, we find finally:
\begin{eqnarray}
  \nonumber M_{dis}^{\bot(2)}(\dtt, B) &\propto&\left(1-\exp\left(-\frac{\dw}{\ts}\right)
  \right)\frac{I_{p,dc}}{\lam}\frac{1}{\Gamma^2+1} \times \\ \nonumber
  & &\times \left\{\exp\left(-\frac{\dtt}{\ts}\right)-\exp \left(-\frac{\dtt}{\tstern} \right)\left[\cos(\lam \dtt)+\Gamma \sin(\lam
  \dtt)\right]\right\}\\
  & & \label{eq:summand2endform}.
\end{eqnarray}

\noindent All the $B$-field dependence of $M_{dis}^{\bot(2)}(\dtt, B)$ is given by $\lam$. The proportionality factor is still independent of $B$. Strikingly, the evolution of the calculated net magnetization in Eq. \ref{eq:summand2endform} is the sum of an
exponentially decreasing background with the characteristic time
constant $\ts$ of the Schottky contact:
\begin{eqnarray} \label{eq:background2}
  M_{bg}^{\bot(2)}(\dtt,B) &=& A_{bg} \exp\left(-\frac{\dtt}{\ts}\right) \\
  A_{bg}&\propto&\left(1-\exp\left(-\frac{\dw}{\ts}\right)
  \right)\frac{I_{p,dc}}{\lam}\frac{1}{\Gamma^2+1} \label{eq:background}.
\end{eqnarray}

\noindent and an exponentially damped oscillation $M_{osc}^{\bot(2)}(\dtt,
B)$ with the time constant $\tstern$ of the spins:

\begin{eqnarray}
  \nonumber M_{osc}^{\bot(2)}(\dtt, B) &\propto&\left(1-\exp\left(-\frac{\dw}{\ts}\right)
  \right)\frac{I_{p,dc}}{\lam}\frac{1}{\Gamma^2+1} \times \\
  & &\times \left\{-\exp \left(-\frac{\dtt}{\tstern} \right)\left[\cos(\lam \dtt)+\Gamma \sin(\lam \dtt)\right]\right\}
\end{eqnarray}

\noindent The latter can be expressed in terms of a net
magnetization of a purely coherently injected spin packet
$M_0^{\bot}(\dt,B)$ (Eq. \ref{eq:singlespin}) with an additional
phase $\delta_2$:

\begin{equation}
  M_{osc}^{\bot(2)}(\dtt, B)= A^{(2)} \exp \left( -\frac{\dtt}{\tstern} \right) \sin
    \left( \omega_L \dtt + \delta_2 \right), \label{eq:oscillation2}
\end{equation}

\noindent with the definitions

\begin{eqnarray}
  \label{eq:amplitude2} A^{(2)} &\propto&  \left(1-\exp\left(-\frac{\dw}{\ts}\right)
  \right)\frac{I_{p,dc}}{|\lam|}\frac{1}{\sqrt{\Gamma^2+1}}\\
  \label{eq:phase2} \tan \delta_2 &=& 1/\Gamma \\
  \dtt &=& \dt-\dw,
\end{eqnarray}

\noindent where we used $\sin(\arctan(1/\Gamma))=1/\sqrt{\Gamma^2+1}$ and $\cos(\arctan(1/\Gamma))=\Gamma/\sqrt{\Gamma^2+1}$.
Note that $\lam<0$ for an effective g-factor $g<0$ as it
is the case for GaAs. Remarkably, the amplitude $A^{(2)}$ of the
precessing net magnetization becomes a function of the absolute
magnetic field $|B|$ because of $|\lam|$ and $\Gamma^2(\lam)$ (see
Eq. \ref{eq:charconstant}). For a vanishing Schottky capacitance
$\ts \rightarrow 0$, which yields a square-like pulsed $I_p(t)$, the
summand $M_{osc}^{\bot(2)}$ vanishes due to $\Gamma \rightarrow
\infty$ and $A^{(2)} \rightarrow 0$. This limit confirms the
interpretation of $M_{osc}^{\bot(2)}$. A huge time constant $\ts
\rightarrow \infty$ suppresses the spin injection $I_p(t)
\rightarrow 0$ (Eq. \ref{eq:polarizedcurrent}) and results consistently in $A^{(2)}
\rightarrow 0$ due to the prefactor $\left(1-\exp\ (-\dw/\ts)
\right)$ in Eq. \ref{eq:amplitude2}.

Finally, we consider the first summand $M_{dis}^{(1)}(\dt,B)$ of Eq.
\ref{eq:term1}. This can be expressed as

\begin{eqnarray}
    \nonumber M_{dis}^{(1)}(\dt,B)&=&c(B) M_0(\dt,B)\\
    &=&c(B) \exp \left( -\frac{\Delta t}{\tstern} \right) \exp
    \left(i \lam \dt \right) \\
    \label{eq:imsummand1} M_{dis}^{\bot(1)}(\dt,B)&=& \Im\left( M_{dis}^{(1)}(\dt,B)
    \right) \\
    &=& A^{(1)} \exp \left( -\frac{\Delta t}{\tstern} \right) \sin \left(\lam \Delta t + \delta_1 \right)
\end{eqnarray}

\noindent with a complex constant $c(B)$, which is the result of the
integral in Eq. \ref{eq:term1} and does not depend on $\dt$ but on
$\dw$. To put it more clearly, the first summand of the net
magnetization during the discharge of the capacitance $C_s$
($\dt>\dw$) in Eq. \ref{eq:term1} is a Larmor precession with frequency $\lam$, decay time
$\tstern$ starting with a phase $\delta_1$. The superposition of the
exponentially damped oscillations $M_{dis}^{\bot(1)}(\dt,B)$ and
$M_{osc}^{\bot(2)}(\dt,B)$ from Eq. \ref{eq:oscillation2} yield a new oscillation with amplitude
$A(B)$ and phase $\delta(B)$. Regarding the exponential background
(Eq. \ref{eq:background}), the measured $\theta_F(\Delta t, B)
\propto \mathfrak{Im}(M(\Delta t, B))$ is thus equivalent to the
fitting formula Eq. 1 of the main article for the considered polarized current $I_p(t)$.

If the voltage pulse repetition time $T_{rep}$ is shorter than the
spin dephasing time $T_2^*$ (see Fig. 4 of the main article), interference of subsequent voltage
pulses has to be taken into account. In this case of resonant spin
amplification, the ansatz Eq. 3 of the article has to be replaced
by Eq. 4. The summation of the voltage pulses leads to resonant spin
amplification and more complicated dependence of the amplitude and
the phase of the oscillating net magnetization $M_{RSA}$ upon application of the
transverse magnetic field. It is not surprising, that resonant spin amplification can also be observed, if the period of the probe $T_{rep, probe}$ is much larger than $\tstern$, but the period of the pump $T_{rep, pump}$ is smaller than $\tstern$ and fulfills the resonance condition $\frac{2 \pi}{T_{rep, pump}} \approx \omega_L$ as shown in section S1 of the supplements. Note that the small positive shift in Faraday rotation $\theta_F$ in Fig. \ref{figs3} originates from the effective non-oscillating background in Eq. \ref{eq:background}. In fact, we can derive the magnetization dynamics of the resonant spin amplification case $M_{RSA}(\dt, B) $analogously to the case of a single pump pulse shown here, but have to solve for $d$ by the condition $I_p(0)=I_p(T_{rep})=I_p(T_{rep, pump})$ in Eq. \ref{eq:polarizedcurrent}:
\begin{equation}
    d=\frac{\exp(\frac{\dw-T_{rep}}{\ts})-1}{\exp(-\frac{T_{rep}}{\ts})-1}.
\end{equation}
\noindent The result was used for the simulations shown in Fig. 4b of the main article.

\section{Magnetic field dependence of the background}

\begin{figure}[tbh]
\includegraphics{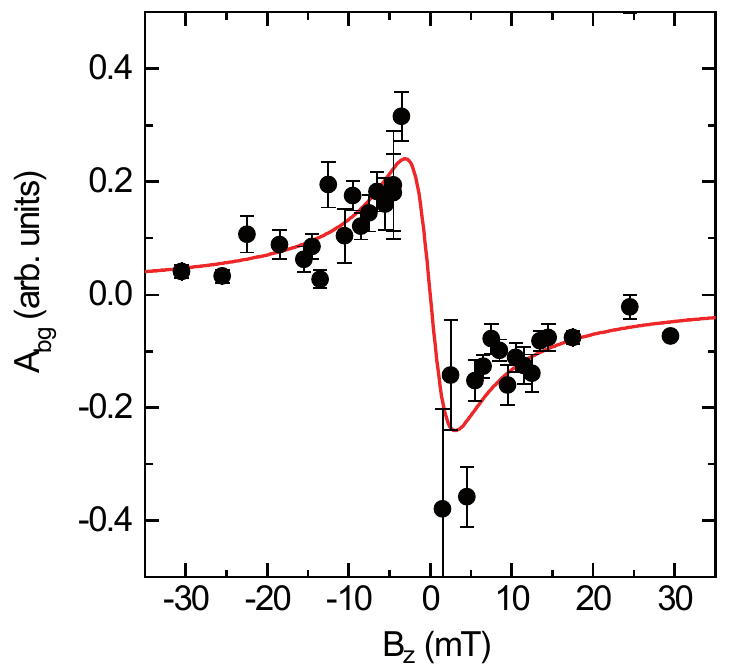}
\caption{\label{figs4} \textbf{Magnetic field dependence of the background signal during pulsed spin injection.} The parameter $A_{bg}$, determined from least-square fits of the measured Faraday rotation curves plotted in Fig. 2a of the main text is plotted as a function of the external magnetic field $B_z$. The error bars include the least-squares fit errors only. The red line represents the expected dependency with the determined parameters according to Eq. \ref{eq:background}.}
\end{figure}

In the main article, the magnetic field dependence of the amplitude $A(B_z)$ of the oscillating component observed during time-resolved spin injection is compared to expectations from our model (Fig. 2d). Here, we discuss the magnetic-field dependence of the amplitude of the non-oscillating background $A_{bg}(B_z)$ (shown in Fig. \ref{figs4}), which exponentially decays as a function $\dt$ as observed in Fig. 2a. Applying Eq.~1, $A_{bg}(B_z)$ is extracted from the least-square fits of the $\theta_F(\dt)$ shown in Fig. 2a, of which the fitted parameters $|g|=0.42$ and $\tau_{bg}=8$\,ns as well as $\ts(B_z)$ and $A(B_z)$ were already discussed in the main article.

According to the model described in Section S2, $A_{bg}(B_z)$ is given by Eq. \ref{eq:background} and parameterized by $|g|$, $\Delta w$, $\tau_{sch}$ and the magnetic field dependence of $\ts(B_z)$. We observe a good agreement of the expected curve with the fitted black data points (Fig. \ref{figs4}), if we use the same values for all the parameters $|g|=0.42$, $\Delta w=2$\,ns, $\tau_{sch} \approx \tau_{bg}=8$\,ns and $\ts(B_z)$ as extracted from the oscillating magnetic field component during time-resolved spin injection. The shape of $A_{bg}(B_z)$ resembles a Hanle depolarization curve, but does not originate from continuous spin injection as in Fig. 1 of the main article. The origin here is that the current pulse triggers a magnetization, a part of which ($M_{dis}(\dt, B)$) can be written as a non-oscillating $M_{bg}^{\bot(2)}(\dt, B)$ (Eq. \ref{eq:background2}) and an oscillating $M_{osc}^{\bot(2)}(\dt, B)$ (Eq. \ref{eq:oscillation2}) component.

\section{Time domain reflectometry}

\begin{figure}[tbh]
\includegraphics{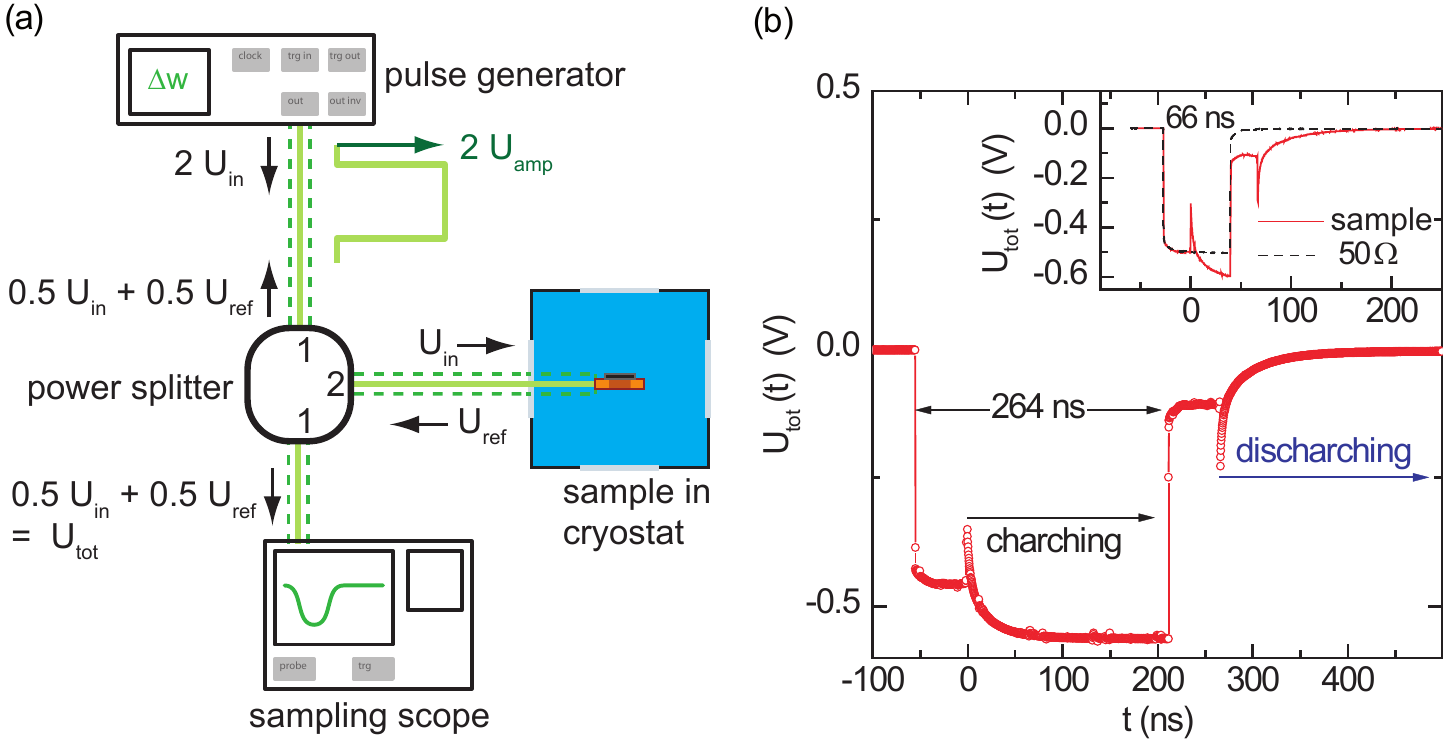}
\caption{\label{figs5} \textbf{Time domain reflectometry setup and measurement. }
(a) Set-up for time domain reflectometry. A pulse pattern generator sends a square pulse through a power splitter to the Fe/GaAs sample placed in the magneto-optical cryostat. The reflected voltage is measured by a sampling oscilloscope in the time-domain.
(b) Time-domain reflectometry performed at 17 K on the Fe/GaAs mesa sample presented in the main article: The voltage $U_{tot}(t)=U_{in}(t)+U_{ref}(t)$ (red circles) is measured as a function of the time $t$ by a sampling oscilloscope. Here, we used a voltage pulse with an amplitude of  $U_{amp} = -0.9$~V and a width of $\Delta w = 264$~ns. $t = 0$~ns correspond to the onset of the reflected signal. The Schottky capacitance starts to discharge at $t = 264$~ns. The outcome of the measurement using $U_{amp} = -1.0$~V and $\Delta w = 66$\,ns is plotted in the inset (red solid line). The black dashed line shows the reflected pulse when using a broadband $50~\Omega$ termination replacing the Fe/GaAs sample.}
\end{figure}

For probing the charging dynamics of the Schottky contact discussed in Fig. 3b of the main article, we added a broadband 50\,\% power splitter to the otherwise unchanged setup (Fig. \ref{figs5}a) and recorded the voltage ($U_{ref}$) back-reflected from the sample together with a part of the voltage applied to the sample $U_{in}$ by a fast sampling scope. The observed total  voltage $U_{tot}=U_{in}+U_{ref}$ (Fig. \ref{figs5}b) reveals the evolution of the voltage at the sample starting at $t=0$\,s. As long as the voltage pulse is applied (total duration $\Delta w=264$\,ns) the Schottky capacity charges up till the reflected voltage saturates. Its saturation value would correspond to $-0.9$\,V$=U_{amp}$, if the parallel resistance $R_S$ to the Schottky capacitance was zero (open termination). The intentional drop of the absolute voltage at $t=0$ can be understood by charging up the fully uncharged Schottky capacitance. Note that a total voltage drop to zero is expected, if the parallel resistance $R_S$ to the Schottky capacitance is zero (shorted termination). The discharging of the Schottky capacitance starting at $t=264$\,ns results in the reversed dynamics. In the inset of Fig. \ref{figs5}b, we compare time-domain reflectometry (here we used a pulse of $U_{amp}=-1$\,V and $\Delta w = 66$\,ns) applied to the sample and to a broadband 50\,Ohm impedance replacing the sample. In the latter case only the $U_{in}$ part of $U_{tot}$ is measured as expected.

\end{document}